\newcounter{myctr}

\documentclass{ws-acs}

\begin{document}

\markboth{Dami\'an H. Zanette}{Zipf's law and city sizes:
Multiplicative processes in urban growth}

\catchline{}{}{}{}{}

\title{ZIPF'S LAW AND CITY SIZES: A SHORT TUTORIAL REVIEW \\ ON
MULTIPLICATIVE PROCESSES IN URBAN GROWTH}

\author{\footnotesize DAMI\'AN H. ZANETTE}

\address{Consejo Nacional de Investigaciones Cient\'{\i}ficas y T\'ecnicas\\
Centro At\'omico Bariloche and Instituto Balseiro \\ 8400 Bariloche,
R\'{\i}o Negro, Argentina
\\
zanette@cab.cnea.gov.ar}

\maketitle

\begin{history}
\received{(received date)}
\revised{(revised date)}
\end{history}

\begin{abstract}
We address the role of multiplicative stochastic processes in
modeling the occurrence of power-law city size distributions. As an
explanation of the result of Zipf's rank analysis, Simon's model is
presented in a mathematically elementary way, with a thorough
discussion of the involved hypotheses. Emphasis is put on the
flexibility of the model, as to its possible extensions and the
relaxation of some strong assumptions. We point out some open
problems regarding the prediction of the detailed shape of Zipf's
rank plots, which may be tackled by means of such extensions.
\end{abstract}


\section{Introduction}

Biological populations --and, among them, human communities-- are
subject, during their existence, to a multitude of actions of quite
disparate origins. Such actions involve a complex interplay between
factors endogenous to the population and external effects related to
the interaction with the ecosystem and with physical environmental
factors. The underlying mechanism governing the growth or decline of
the population size (i.e., the number of individuals) is however
very simple in essence, since it derives from the elementary events
of reproduction: at a given time, the growth rate of the population
is proportional to the population itself. This statement must be
understood in the sense that two populations formed by the same
organisms and under the same ecological conditions, one of them
--say-- twice as large as the other, will grow by amounts also
related by a factor of two. Such proportionality between population
and growth rate, which is empirically verified in practically all
instances of biological systems, defines a {\it multiplicative
process} \cite{Sornette2}.

Populations whose size is governed by multiplicative processes and
which, at the same time, are subject to environmental random-like
fluctuations, are known to display universal statistical
regularities in the distribution of certain features. Specifically,
those traits which are transmitted vertically, from parents to their
offspring, exhibit broad, long-tailed distributions with stereotyped
shapes --typically, log-normal or power laws.  For instance,
consider a human society where, except for some unfrequent
exceptions, the surname of each individual is inherited from the
father. Consider moreover the subpopulations formed by individuals
with the same surname. It turns out that the frequency of
subpopulations of size $n$ is approximately proportional to $n^{-2}$
\cite{Z0,Man:03}. Or take, from the whole human population, the
communities whose individuals speak the same language, which in the
vast majority of the cases is learnt from the mother. The sizes of
those communities are distributed following a log-normal function
\cite{lang}. Such statistical regularities are generally referred to
as {\it Zipf's law} \cite{Zipf1,Z0}. The derivation of Zipf's law
from the underlying multiplicative processes was first worked out in
detail by the sociologist H. A. Simon, within a set of assumptions
which became known as {\it Simon's model} \cite{Simon}.

A well-documented instance of occurrence of Zipf's law involves the
distribution of city sizes \cite{Gibrat,nos,Gabaix2,Gabaix1}, where
``size'' is here identified with the number of inhabitants. In
practically any country or region over the globe, the frequency of
cities of size $n$ decays as $n^{-z}$, where the exponent $z$ is
approximately equal to $2$ --as in the case of surnames. The
occurrence of Zipf's law in the distribution of city sizes can be
understood in terms of multiplicative processes using Simon's model.
Inspection of current literature on the subject of city size
distributions, however, suggests that the potential of Simon's model
as an explanation of Zipf's law, as well as its limitations, are not
well understood. In a recently published handbook on urban economics
\cite{Gabaix1}, for instance, we read: {\it ``Simon's model
encounters some serious problems. In the limit where it can generate
Zipf's law, it ... requires that the number of cities grow
indefinitely, in fact as fast as the urban population.''} It turns
out that this assertion is wrong: the truth, in fact, happens to be
exactly the opposite! Leaving aside the derivation that may have led
to this false conclusion \cite{Gabaix2}, we note that such strong
statements risk  to become dogmatic for the part of the scientific
community which does not have the tools for their critical analysis.

With this motivation, the present short review will be devoted to
give a pedagogical presentation of Simon's model in the frame of the
evolution of city size distributions. The emphasis will be put on a
qualitative description of the basic processes involved in the
modeling. The explicit statement of the hypotheses that define the
model should already expose its limitations but, at the same time,
should clarify its flexibility regarding possible generalizations.
In the next section,  an elementary model for the evolution of a
population based on stochastic processes is introduced, and the
concurrent role of multiplicative and additive mechanisms in the
appearance of power-law distributions is discussed. After an outline
of the main features of Zipf's rank plots in the distribution of
city sizes, Simon's model is presented in its original version,
describing its implications as for the population distribution in
urban systems. Then, we discuss a few extensions of the model, aimed
at capturing some relevant processes not present in its original
formulation. Finally, we close with a summary of the main results
and some concluding remarks.

\section{Multiplicative processes and population growth \label{multi}}

The fluctuating nature of the many environmental actions which
modulate the growth of a population calls for a description based on
stochastic --i.e., random-- processes. Within this kind of
formulation, it is explicitly assumed that the parameters that
govern the evolution  can change with time in irregular ways. For
instance, the change in the number $n(t)$ of individuals during a
certain time interval $\Delta t$ can be modeled by means of the
discrete stochastic equation
\begin{equation} \label{mult}
n(t+\Delta t)- n(t) = a(t) n(t) +  f(t)
\end{equation}
where $a(t)$ and $f(t)$ are random variables. At each time step,
their values are drawn from suitably chosen probability
distributions. As a consequence of the random variation of $a(t)$
and $f(t)$, the number $n(t)$ also displays stochastic evolution.
Equation (\ref{mult}) is used to predict the statistical properties
of $n(t)$, for instance, finding the probability distribution
$P(n,t)$ that the population has a value $n$ a time $t$. This kind
of equation has been studied in detail by several authors in various
contexts \cite{Sornette2,Sornette1}.

The first term in the right-hand side of Eq. (\ref{mult}),
$a(t)n(t)$, represents the contributions to the evolution of $n$
that are proportional to the population itself, i.e. the {\it
multiplicative} effects referred to in the Introduction. If the
population is closed, multiplicative processes are restricted to
birth and death, and $a(t)$ stands for the difference between the
birth and death rates per individual in the interval $\Delta t$. In
open populations, the number of individuals is also affected by
migration processes. Emigration flows are generally proportional to
$n(t)$ because, on the average, each individual has a certain
probability of leaving the population per time unit. On the other
hand, immigration has both multiplicative and {\it additive}
effects. In fact, immigration can be favored by a large preexisting
population --as in big cities-- but a portion of arrivals may also
occur as a consequence of individual decisions that do not take into
account how large the population is. Additive contributions are
described by the second term in Eq. (\ref{mult}). This term can also
stand for negative effects on the population growth, such as
catastrophic events where a substantial part of the population dies
irrespectively of the value of $n(t)$ \cite{reset}.

The probability distribution $P(n,t)$ of the population $n$, as
derived from Eq. (\ref{mult}), can have a complicated analytical
form depending on the specific distributions chosen for $a(t)$ and
$f(t)$. It is nevertheless known that, for large times, it decreases
as a power law,
\begin{equation}  \label{powerlaw}
P(n,t) \sim n^{-1-\gamma},
\end{equation}
over a substantial interval of values of $n$. The exponent
$\gamma>0$ is given by the distribution $p(a)$ for the random
variable $a(t)$, as the solution of the equation \cite{Sornette1}
\begin{equation}
1 = \int p(a) (a+1)^\gamma da.
\end{equation}
Equation (\ref{powerlaw}) holds under very general conditions on the
probability distributions of $a(t)$ and $f(t)$, provided that $f(t)$
is not identically equal to zero. In other words, a power-law
distribution is obtained when both multiplicative and additive
processes are in action. The case $f(t)\equiv 0$ is, mathematically
speaking, a singular limit. In the absence of additive processes,
$P(n,t)$ becomes a log-normal distribution.

The empirical observation of a power-law distribution in a real
system would require to have access to many realizations of the
evolution of the same population --which, in practice, is rarely
possible-- or, alternatively, to follow the parallel evolution of
several sub-populations of the same type. This second instance is
often met in human populations, which are naturally divided into
communities of different nature, determined by historical,
geographical, sociocultural, and/or economic factors. One of such
divisions is given, precisely, by urban settlements. In this case,
$P(n,t)$ can be interpreted as the probability of having a city of
size $n$ at time $t$ within the region that encompasses the whole
population under study. In view of the above discussion, it is
expected that the populations of different cities follow, under
suitably homogeneous conditions over the studied region, a power-law
distribution. As is well known, in fact, they do. Power laws in the
population distribution of human groups of various kinds have been
reported by several authors and, notably, by the philologist  G. K.
Zipf \cite{Z0}. As advanced in the Introduction, the power-law
dependence of the frequency of groups as a function of their
population is now known as Zipf's law.

Zipf's law is often presented in an alternative formulation which,
in the frame of the distribution of city sizes, goes as follows.
Take all the cities under consideration, and rank them in order of
decreasing population, so that rank $r=1$ corresponds to the largest
city, $r=2$ to the second largest, and so on. Then, the population
$n$ of a city decreases with its rank as a power law,
\begin{equation} \label{Zipf}
n(r) \sim r^{-z} ,
\end{equation}
over a wide range of values of $r$. The exponent $z$ is usually
referred to as the Zipf exponent. Equations (\ref{powerlaw}) and
(\ref{Zipf}) are closely related. In fact, the formulation of Zipf's
law in terms of the probability distribution $P(n,t)$ and the rank
formulation are equivalent, though the latter is much less
significant than the former from the viewpoint of a statistical
description. To understand the connection between the two
formulations, it is first useful to recall that --in our
interpretation of the stochastic growth equation (\ref{mult}) as
describing the parallel evolution of several sub-populations-- the
sum
$$
\sum_{n=n_1}^{n_2} P(n,t)
$$
is the probability of having a city with population between $n_1$
and $n_2$. Accordingly, the product of this sum times the total
number of cities under consideration, is the number of cities with
populations within that interval. Since the rank $r$ of a city of
population $n$ equals the number of cities with populations larger
than or equal to $n$, we have
\begin{equation} \label{r(n)}
r(n) = M \sum_{n'=n}^\infty P(n',t) dn',
\end{equation}
where $M$ is the total number of cities. This establishes the
relation between the rank-frequency dependence and the probability
distribution $P$. In particular, replacing Eq. (\ref{powerlaw}) into
(\ref{r(n)}), we get $r(n) \sim n^{-\gamma}$, implying that the Zipf
exponent and $\gamma$ are related as
\begin{equation} \label{zgamma}
z=\frac{1}{\gamma} .
\end{equation}
This defines the connection between the power laws in Eqs.
(\ref{powerlaw}) and (\ref{Zipf}).

\begin{figure}[th]
\centerline{\psfig{file=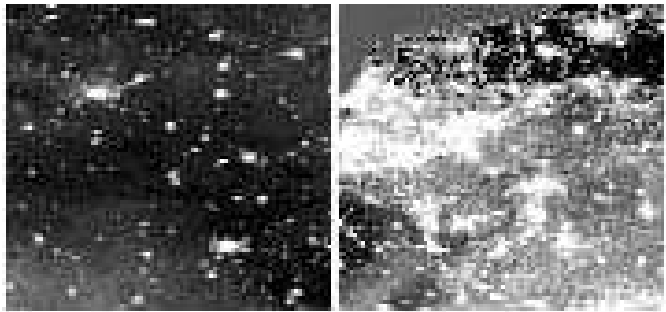,width=3.5in,clip=}} \caption{Two
satellite images of the Earth by night. Left: Central Ukraine.
Right: North-western Germany. Each image covers an area of, roughly,
$500  \times 500$ km$^2$. Source: visibleearth.nasa.gov. \label{gu}}
\end{figure}

\section{Zipf's law in the distribution of city sizes} \label{z}

\begin{figure}[th]
\centerline{\psfig{file=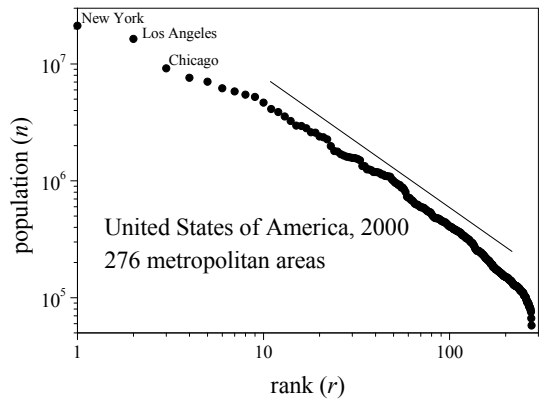,width=3in,clip=}} \caption{Zipf
rank plot for $276$ metropolitan areas in the United States, after
results of the census in $2000$. Source: factfinder.census.gov. The
straight line has slope $1.11$. \label{zipfusa}}
\end{figure}

The application of Zipf's rank analysis to urban settlements
implicitly assumes that individual cities are well-defined entities.
Actually, however, the modern city is such a complex of intermingled
systems that it defies a definition in terms of traditional
classification schemes, and requires a wider concept of class
\cite{Portu}. Figure \ref{gu} illustrates the fact that, while
individual urban settlements can be distinctly identified in some
regions, in other places the situation is by far less obvious.
Anyway, it is currently accepted that the entities to be considered
in Zipf's analysis are the clusters resulting from the growth and
aggregation of initially separated settlements. A plot of the
population $n$ versus the rank $r$ for the cities of a given country
or region usually reveals three regimes. For the lowest ranks,
corresponding to the largest cities, the variation of $n$ with $r$
is generally irregular, with a marked descending step between the
first one to three cities and the following. The biggest urban
settlements in any large country or region often lie outside any
significant statistical regularity, both within the region in
question and between different regions. As the rank becomes higher,
these irregularities smooth out, and the plot enters the power-law
regime. In the usual representation of $n$ versus $r$ in log-log
scales, this regime is revealed by a linear profile, typically
extending from $r \approx 10$ to ranks of the order of a few to
several hundreds. The Zipf exponent $z$, given by the slope of the
linear profile in the log-log plot, is considerably uniform between
different regions. It is customary to quote the value $z\approx 1$,
though it may vary between $0.7$ and $1.2$. Finally, for the highest
ranks the power-law regime is cut off, and $n$ declines faster as
$r$ grows. Figure \ref{zipfusa} illustrates these typical features
for $276$ metropolitan areas in the USA, according to data from the
year $2000$ census. Note carefully that the class of ``metropolitan
areas'' does not necessarily include all urban settlements above a
certain size. Below, we comment on related methodological problems
in the construction of rank plots.

It is clear from the discussion on multiplicative stochastic
processes in Section \ref{multi} that, among the three regimes
identified in rank plots, the natural candidate to be explained in
terms of such mechanism is the central power-law range. It also
results from our discussion that, to derive a power-law city size
distribution, it is necessary to take into account both
multiplicative and additive contributions to the evolution of the
population. These ingredients are captured by Simon's model, which
is presented in next section. Let us here point out that, to explain
Zipf's law in the distribution of city sizes, a model solely based
on multiplicative processes --namely, Gibrat's model \cite{Gibrat}--
is often invoked. As already commented, however, purely
multiplicative mechanisms can only produce a log-normal
distribution. While over restricted ranges a log-normal distribution
may seem to exhibit a power-law decay, $P(n,t) \sim n^{-\lambda}$
with $\lambda \approx 1$, it certainly cannot fit the variety of
Zipf exponents found in real city size distributions.

Before passing to the formulation of Simon's model for the power-law
regime of rank plots, it is pertinent to discuss a few aspects
regarding the description of the two remaining regimes --those
corresponding to the lowest and the highest ranks. As for the
former, the biggest cities in a large country or region are, almost
invariably, special cases that elude inclusion in any statistical
description. It would be hopeless to pretend that, for instance,
Paris, Berlin, or Rome enter the same statistical class as the
European cities whose present population is below, say, one million.
The political and economic role of those cities has been --and still
is-- markedly peculiar. In consequence, their individual evolution
is exceptional among urban settlements and must be dealt with as
such. While it would make little sense to discuss the case of the
biggest cities in the frame of a statistical model for the
distribution of city sizes, it is nevertheless interesting to
advance that Simon's model assigns a special role to those cities:
their sizes bear information on the initial state of the urban
system, before the smaller settlements played any significant role
in the population statistics. In this sense, Simon's model also
recognizes that the biggest cities are special cases.

As for the cut-off region of highest ranks, let us mention that it
is found not only in rank plots for city sizes, but also in many
other instances where Zipf's law holds for intermediate ranks. A
classical example occurs in the frequency of words in human
languages \cite{MM1,MM2}. In the case of city sizes, the appearance
of the cut-off is well known but, to our knowledge, there is no
systematic study regarding the population-rank functional dependence
in that regime. This lack of quantitative empirical results
discourages modelling of city sizes for high ranks, as there is no
reference data to validate potential models. Moreover, as pointed
out in connection with Fig. \ref{zipfusa}, the regime of high ranks
is susceptible of methodological errors related to possible data
incompleteness. While, arguably, the lists of large cities provided
by most sources of demographic information are exhaustive, the same
sources may result to be less reliable when it comes to smaller
urban settlements. Inspection of many public-domain databases
immediately reveals lack of completeness in the lists of cities for
high ranks. The direct effect of these ``gaps'' is that the assigned
ranks are lower than in reality, with the consequent reinforcement
of the cut-off (cf. Fig. \ref{zipfusa}). Avoiding this effect
without restricting too much the range of ranks under consideration,
requires relying on presumably complete data sets --typically, from
official census reports. This, in turn, limits the corpus of data,
because such databases are not always available. In any case, as
stated above, the cut-off in rank plots can be observed in other
systems where this kind of methodological error is not present. In
Section \ref{extens}, we adapt to the case of city sizes an
extension of Simon's model put forward to give a semi-quantitative
explanation of the cut-off regime for the case of word frequencies
in language.

\section{Simon's model: Hypotheses and main results \label{Simonmod}}

Elaborating on an idea previously advanced by Willis and Yule
\cite{Yule}, H. A. Simon proposed in $1955$  the model that now bear
his name \cite{Simon}, as an explanation for the origin of power-law
distributions and Zipf's law. Simon presented his model by referring
to the case of word frequencies, which Zipf himself had discussed in
detail in his publications \cite{Zipf1}. Here, we  introduce the
original Simon's model adapted to the framework of city growth. In
practice, this just implies a change in the vocabulary employed to
express the dynamical rules that define the model.

Simon's model describes the evolution of a population divided into
well-defined groups --the cities. We characterize this division by
means of the quantity $m(n,t)$, which gives the number of cities
with exactly $n$ inhabitants at time $t$. This quantity is closely
related to the probability distribution $P(n,t)$ introduced in
Section \ref{multi}. In fact, we have $m(n,t)=M(t) P(n,t)$, where
$M(t)$ is the total number of cities in the system at the same time.
Instead of using the real time $t$, Simon's model proceeds by
discrete steps, which are identified by means of a discrete variable
$s=0,1,2\dots$ Each step corresponds to the time interval needed for
the total population to increase by exactly one inhabitant. The
actual duration of an evolution step --which is determined  by a
balance between birth and immigration on one side, and death and
emigration on the other-- is irrelevant to the model. The growth of
the population in real time is a separate problem which can be
specified and solved independently. As for the model, thus, the
elementary evolution event is the addition of a single inhabitant to
the total population. Accordingly, the quantity $m(n,t)$ is replaced
by $m(n,s)$, the number of cities with exactly $n$ inhabitants at
step $s$. The starting point of the evolution is given by the
initial distribution $m(n,0)$, at $s=0$, which describes a
preexistent population distributed over a certain number of cities.

The evolution is governed by the following stochastic rules, which
imply making a decision at each step $s$, i.e. each time a new
inhabitant is added to the total population. (i) With probability
$\alpha$, the new inhabitant founds a new city. In this case, the
number of cities $M$ grows by one, and the new city has initially a
single inhabitant. (ii) With the complementary probability,
$1-\alpha$, the new inhabitant is added to an already populated
city. In this case, the destination city is chosen with a
probability proportional to its current population. It increases its
population by one, and the number of cities $M$ does not vary.
Clearly, rule (ii) stands for the multiplicative contribution to the
evolution of the individual population of cities. Larger cities have
higher probability of incorporating new inhabitants than smaller
ones. As it grows, the population is preferentially assigned to
those groups which are already relatively large. Rule (i), on the
other hand, represents a contribution independent of the preexisting
distribution and, thus, stands for additive effects. In particular,
it implies that the number of cities grows, on the average, at a
constant rate $\alpha$. Hence, the average number of cities at step
$s$ is $M(s)=M(0)+\alpha s$. Meanwhile, since exactly one inhabitant
is added per time step, the  total population in the system is
$N(s)=N(0)+s$.

In order to translate into mathematical terms the evolution rules
(i) and (ii), we must take into account the following remarks.
First, rule (i) only affects the number of cities with exactly one
inhabitant, $m(1,s)$. When it applies, which happens with frequency
$\alpha$ per evolution step, $m(1,s)$ grows by one. Second, when
rule (ii) applies, which happens with frequency $1-\alpha$ per
evolution step, the probability that any city of size $n$ is chosen
as destination is proportional to $n/N(s)$ and to the number of
cities of that size. Since the chosen city changes its population
from $n$ to $n+1$, this event represents a positive contribution to
the number of cities of size $n+1$ at the next step, $m(n+1,s+1)$,
and a negative contribution to $m(n,s+1)$. The contribution to
$m(n,s+1)$ will be positive when the chosen destination is a city of
size $n-1$. Summing up these considerations, the average change per
step in the number of cities of size $n$ is
\begin{equation} \label{Simon1}
m(1,s+1)-m(1,s)=\alpha-\frac{1-\alpha}{N(s)} m(1,s)
\end{equation}
for $n=1$, and
\begin{equation} \label{Simonn}
m(n,s+1)-m(n,s)=(1-\alpha)\left[ \frac{n-1}{N(s)}
m(n-1,s)-\frac{n}{N(s)} m(n,s) \right]
\end{equation}
for $n=2,3,4\dots$ These are the equations that govern the evolution
of Simon's model in its original formulation \cite{Simon}.

From a mathematical viewpoint, Eqs. (\ref{Simon1}) and
(\ref{Simonn}) are not complicated. First of all, they form a linear
system, which can therefore be tackled with a host of well-tested
analytical and numerical methods. Moreover, they can be solved
recursively. In fact, the solution to Eq. (\ref{Simon1}) gives the
number of cities with one inhabitant. Once $m(1,s)$ has been found,
$m(2,s)$ and, successively, $m(n,s)$ for larger $n$, are obtained
from Eq. (\ref{Simonn}). The only difficulty is that the equations
involve the function $N(s)=N(0)+s$, which depends explicitly on the
variable $s$. Consequently, the system is non-autonomous.

In his original paper, Simon was able to prove that --as we show
below-- Eqs. (\ref{Simon1}) and (\ref{Simonn}) imply a power-law
decay for $m(n,s)$ as a function of $n$. The presentation of the
solution will differ from Simon's in that we first introduce a
continuous approximation to the model equations, replacing the
discrete variables $n$ and $s$ by continuous variables $\eta$ and
$\xi$, respectively. This approximation has the advantage of
transforming the infinitely many equations (\ref{Simonn}) into a
single evolution law. The disadvantage is that the new problem is
differential, instead of algebraic. Replacing discrete by continuous
variables is justified by the fact that, in the distribution of city
sizes, we are mainly interested in the range of large values for
both $n$ and $s$, where $m(n,s)$ is expected to vary smoothly. To
the first order, we approximate the differences in Eq.
(\ref{Simonn}) by derivatives; for instance,
$$
m(n,s+1)-m(n,s) \approx \frac{\partial m}{\partial \xi} (\eta,\xi).
$$
This approximation can be systematically improved by considering
higher order terms, as discussed elsewhere \cite{MZJTB}. The
resulting partial-differential equation is
\begin{equation} \label{conti}
\frac{\partial \mu}{ \partial \xi} +(1-\alpha) \frac{\eta}{N(0)+\xi}
\frac{\partial \mu}{ \partial \eta} =0,
\end{equation}
with $\mu ( \eta,\xi)=\eta m(\eta,\xi)$. This equation has to be
solved for $\eta>1$, with the initial condition $\mu (\eta,0) = \eta
m(\eta,0)$ and a boundary condition at $\eta=1$ derived from Eq.
(\ref{Simon1}), namely,
\begin{equation}
\mu(1,\xi)=\frac{\alpha}{1-\alpha} (N(0)+\xi) .
\end{equation}
We do not discuss the details of the solution method for this linear
equation. It is  enough to say that, by means of a change of
variables \cite{MZJTB}, the equation reduces to a standard
one-dimensional wave equation and is then solved by the so-called
``method of characteristics.'' In the following, we describe the
result in terms of the original variables $n$ and $s$.

As a function of the population $n$, the number $m(n,s)$ of cities
of size $n$ at step $s$, solution of the Simon's equations, shows
two distinct regimes. The boundary between both regimes is situated
at
\begin{equation}
n_B (s) = \left( 1+ \frac{s}{N(0)}\right)^{1-\alpha}.
\end{equation}
We see that this boundary depends on $s$, and shifts to higher
populations as the evolution proceeds. For populations above the
boundary, $n>n_B$, the solution to the continuous approximation of
Simon's model is
\begin{equation}
m(n,s)= \frac{1}{n_B(s)} m\left( \frac{n}{n_B(s)},0 \right).
\end{equation}
Thus, $m(n,s)$ is directly given by the initial condition $m(n,0)$.
As a matter of fact, this regime can be seen to encompass those
cities that where already present at $s=0$. In Simon's model,
preexisting cities --not unlike the oldest cities of real urban
systems--  are those that reach the largest sizes, i.e. those that
are assigned the lowest rank values in Zipf's analysis. We realize
that, as advanced in Section \ref{z}, information about the initial
state of the urban system is stored in the size distribution at the
lowest ranks. A detailed study of the effects of the initial
condition in the large-size regime, referring  to the discrete
equations (\ref{Simon1}) and (\ref{Simonn}), has been presented
elsewhere \cite{PhysA}.

In the range of small populations, $n<n_B$, the solution to the
continuous approximation of Simon's model is
\begin{equation}
m(n,s)= \frac{\alpha}{1-\alpha} (N(0)+s) \, n^{-1-1/(1- \alpha)} .
\end{equation}
This regime encompasses the cities founded during the evolution of
the urban system, corresponding to higher rank values in Zipf's
analysis. We see that their size distribution follows a power law
with exponent $\gamma=1/(1-\alpha)$ [cf. Eq. (\ref{powerlaw})].
According to Eq. (\ref{zgamma}), the Zipf exponent is
\begin{equation}
z=1-\alpha.
\end{equation}
Since, being a probability, $\alpha$ is positive and lower than one,
this formulation of Simon's model predicts a Zipf exponent $0<z<1$.
The characteristic value $z\approx 1$ is obtained for very small
$\alpha$, i.e. {\it when the frequency of city foundation is very
small as compared with the growth rate of the population}. As
advanced in the Introduction, this conclusion is in full
disagreement with the bibliographic quotation given there.

In summary, the main results obtained in this section for the
original version of Simon's model, within the continuous first-order
approximation, are the following. At any evolution stage, the
distribution of city sizes shows two well-differentiated regimes.
For large cities, which correspond to low rank values, the
distribution depends sensibly on the initial condition. This range
keeps information on the early state of the urban system and, thus,
results to be specific for each realization of the model. On the
other hand, the size distribution of small cities, within the range
of high rank values, exhibits a universal power-law decay whose
exponent is completely determined by the rate $\alpha$ at which new
cities are founded. The respective Zipf exponent is always less than
one, and the limit $z=1$ is approached when $\alpha$ is vanishingly
small. The two regimes, whose mutual boundary recedes towards high
populations as the evolution proceeds, can be immediately identified
with two of the three regions of rank plots, described in Section
\ref{z}. The cut-off region, on the other hand, remains unexplained
by this version of Simon's model. Moreover, as presented in this
section, the model is not able to produce Zipf exponents larger than
one (cf. Fig. \ref{zipfusa}). Some of the generalizations discussed
in the next section are aimed at alleviating these limitations.

\section{Generalization of Simon's model \label{extens}}

It is clear that, in the original formulation of Simon's model, both
rules (i) and (ii) involve strong assumptions on the parameters that
govern the evolution of the urban system. Specifically, rule (i)
establishes that the rate at which new cities are founded is
constant, i.e. does not vary with time. Rule (ii), in turn, makes a
concrete hypothesis on the size dependence of the probability for a
city to be chosen as destination for a new inhabitant. Not without
reason, it may be argued that these assumptions are unrealistically
simple. Reinforcing this impression, we have just shown that Simon's
model is not able to predict some basic features in the rank plots
of real urban systems, such as the cut-off at high ranks and the
possibility that the Zipf exponent is larger than one.

It has to be understood, however, that the assumptions implicit in
the evolution rules have been introduced by Simon, mainly, to
facilitate the analytical treatment of the equations and to show, as
straightforwardly as possible, that a couple of elementary
mechanisms are enough to explain the occurrence of power-law
distributions and Zipf's law. If one intends to be more realistic,
those strong assumptions can be immediately relaxed, without
inherently modifying the basic dynamical processes that define the
evolution. In this section, we present a small collection --by no
means exhaustive-- of generalizations of Simon's model, based on
relaxing the evolution rules. Some of these extensions have already
been introduced in the literature to solve the above discussed
limitations of the model, regarding the detailed prediction of
Zipf's rank distributions. Our main aim is, nevertheless, to
emphasize the flexibility of Simon's model as for possible
extensions towards a more realistic description of city growth.

\subsection{Time-dependent rate of city foundation} \label{alfas}

A straightforward generalization of Simon's model consists in
assuming that the probability $\alpha$ of foundation of a city when
a new inhabitant is added to the system depends on time. Indeed, it
is expected that the rate at which new cities appear in a urban
system decreases as the total number of cities grow. In the model, a
time-dependent rate of city foundation amounts at admitting that
$\alpha$ depends on the variable $s$. In this way, $\alpha(s)$ gives
the probability per evolution step that a city is founded by the
inhabitant added at step $s$.

To mathematically implement this generalization, we do not need to
to rewrite the evolution equations. It is just enough to take into
account that, in Eqs. (\ref{Simon1}) and (\ref{Simonn}), the
parameter $\alpha$ may depend on $s$. In principle, there are no
limitations on the functional form of this dependence. Of course,
however, whether the resulting evolution equations are analytically
tractable and whether they produce a power-law distribution is a
matter to be ascertained in each particular case. In any case, the
problem can be dealt with by numerical means.

As an illustration, we consider here a phenomenological model for
the time variation of $\alpha$ put forward in the framework of word
frequencies in language \cite{MM1,MM2}. In this model, $\alpha$
decreases with $s$ as a power law of the form
\begin{equation}
\alpha(s) = \alpha_0 s^{\nu-1},
\end{equation}
where $\alpha_0$ is a constant, and $0<\nu<1$. In the problem of
city growth, this form of $\alpha (s)$ implies that the total number
of cities increases slower than linearly, $M(s) \sim s^\nu$, instead
of displaying linear growth as in the original version of Simon's
model. In the relevant limit where $\alpha (s) \ll 1$ for all $s$
--a condition which is insured if the constant $\alpha_0$ is very
small-- it is possible to find the solution of the first-order
continuous approximation, Eq. (\ref{conti}). As a function of the
population $n$,  the resulting distribution $m(n,s)$ shows again two
regimes. As in the case of constant $\alpha$, the large-population
regime is determined by the initial condition and, thus, bears
information on the initial state of the urban system. The
small-population regime, in turn, corresponds again a power-law
distribution, but its exponent has changed:
\begin{equation}
m(n,s) = \alpha_0 N(0) \left( 1+\frac{s}{N(0)}\right)^\nu
n^{-1-\nu}.
\end{equation}
The associated Zipf exponent is [cf. Eqs. (\ref{powerlaw}) and
(\ref{zgamma})]
\begin{equation}
z=\frac{1}{\nu} .
\end{equation}
Since $\nu <1$, we have $z>1$. We conclude that allowing the rate of
city foundation to depend on time, the restriction in the resulting
Zipf exponent can be removed. The effect of more general forms of
$\alpha(s)$ may be assessed numerically.

\subsection{The cut-off regime}

Another extension of Simon's model makes it possible to predict the
presence of a faster population decay for high ranks, thus providing
a plausible explanation for the cut-off observed in the rank plot.
Here, we limit ourselves to a semi-quantitative description of this
generalization, as technical details have already been given
elsewhere \cite{MM1,MM2}.

The generalization is based on a realistic consideration regarding
the foundation of  cities as new inhabitants are added to the
population. It can be argued that a single inhabitant is not enough
to define the existence of a new city. Rather, there should be a
minimal population for a city to enter the regime where the
multiplicative process of Simon's rule (ii) acts. This effect can be
implemented by modifying the probability that a newly founded city
is chosen as destination by new inhabitants. Namely, the probability
that a city of size $n$ is chosen by a new inhabitant can be taken
to be proportional to $\max \{n, n_{\min}\}$, where $n_{\min}$ is
the threshold population. In this way, a given city must attract
$n_{\min}$ new inhabitants before multiplicative growth begins to
act. Until then, the probability that the city is chosen as
destination is a constant. Note that the threshold $n_{\min}$ may be
different for each city. Within this extension, the cut-off of
Zipf's plot is interpreted as corresponding to those cities whose
size has not yet attained the threshold.

This generalization of Simon's model has originally been introduced
in the framework of word frequencies \cite{MM1}. Numerical
simulations of the model with an exponential distribution for the
value of $n_{\min}$ assigned to each city, combined with an
$s$-dependent probability $\alpha$ of the type discussed in Section
\ref{alfas}, have provided excellent fittings of Zipf rank plots for
several texts in different languages. In view of these encouraging
results, it would be interesting to try these combined extensions of
Simon's model for city size distributions.

\subsection{Size-dependent choice of the destination city}
\label{destin}

As mentioned above, rule (ii) in the original formulation of Simon's
model involves the very special assumption that the probability for
a city to be chosen as destination by a new inhabitant is
proportional to its size. In other words, the specific growth rate
of cities per time step --relative to their current individual
populations-- is constant all over the system.

This assumption can be relaxed supposing that the probability that a
city receives a new inhabitant is not proportional to its population
$n$, but to a function $\phi(n)$. If $\phi(n)$ grows with $n$ faster
than linear, the specific growth rate of large cities will be higher
than for small cities, and {\it vice versa}. In the original
formulation of the model, one has $\phi(n)=n$. The function
$\phi(n)$  stand thus for a nonlinear effect in the multiplicative
process. The probability that a city of size $n$ is chosen as
destination is given by the ratio $\phi(n)/\Phi(s)$, where the
normalization factor is given by
\begin{equation} \label{Phi}
\Phi(s) = \sum_{n=1}^\infty \phi(n) m(n,s).
\end{equation}
This normalization insures that the sum of the probabilities over
the whole ensemble of cities equals one. In the original model, the
normalization factor equals the total population, $\Phi(s)=N(s)$.

Within this generalization, the model evolution equations read
\begin{equation} \label{Simon1a}
m(1,s+1)-m(1,s)=\alpha-(1-\alpha)\frac{\phi(1)}{\Phi(s)} m(1,s)
\end{equation}
for $n=1$, and
\begin{equation} \label{Simonna}
m(n,s+1)-m(n,s)=(1-\alpha)\left[ \frac{\phi(n-1)}{\Phi(s)}
m(n-1,s)-\frac{\phi(n)}{\Phi(s)} m(n,s) \right]
\end{equation}
for $n=2,3,4\dots$ In the first-order continuous approximation
introduced in Section \ref{Simonmod}, they transform into
\begin{equation} \label{contia}
\frac{\partial \psi}{ \partial \xi} +(1-\alpha)
\frac{\phi(\eta)}{\Phi(\xi) } \frac{\partial \psi}{ \partial \eta}
=0,
\end{equation}
with $\psi ( \eta,\xi)=\phi(\eta) m(\eta,\xi)$. As in the original
model, this equation has to be solved for $\eta>1$, with the initial
condition $\psi (\eta,0) = \phi(\eta) m(\eta,0)$. The boundary
condition at $\eta=1$ is now
\begin{equation}
\psi(1,\xi)=\frac{\alpha}{1-\alpha} \phi(1) \Phi(\xi) .
\end{equation}

Now, finally, we have managed to end up with a really complicated
mathematical problem. Equations (\ref{Simon1a}) to (\ref{contia})
are very similar to the evolution equations of the original model
but, alas, the similitude is only formal. The key difficulty of our
new equations for $m(n,s)$ resides in the fact that the function
$\Phi (s)$ is generally not known beforehand. In the original model,
on the other hand, it coincides with the total population
$N(s)=N(0)+s$. Within the present generalization, $\Phi(s)$ can only
be given in terms of $m(n,s)$ itself [cf. Eq. (\ref{Phi})].
Unfortunately, it is not possible to find an independent equation
for the evolution of $\Phi(s)$ alone. The distribution of city sizes
$m(n,s)$ and  $\Phi(s)$ must therefore be found simultaneously and
self-consistently.

We have been unable to find a form of $\phi(n)$ allowing us to give
an analytical solution either to Eqs. (\ref{Simon1a}) and
(\ref{Simonna}) or to Eq. (\ref{contia}). It seems, not
unexpectedly, that the problem must be treated numerically. We leave
it open for the reader interested at studying the effects of
nonlinear multiplicative processes. To our knowledge, this kind of
processes have until now received relatively little attention.

\section{Conclusion}

This short review has been devoted to a presentation of the
mathematics of Simon's model in terms that, we hope, are accessible
to a broad academic readership. We have shown that, in its original
version, Simon's model is able to explain the occurrence of a
power-law regime in the distribution of city sizes, though it fails
at predicting some of the Zipf exponents observed in real urban
systems, as well as other systematic features resulting from Zipf's
rank analysis. The extensions discussed later should have
demonstrated that such limitation can be removed --at least,
partially-- by relaxing some of the assumptions of the model's
dynamical rules without modifying the key underlying mechanisms.
These extensions were mainly aimed at illustrating the potential of
Simon's model with respect to possible generalizations in the
direction of a better description of empirical data.

Several processes relevant to the evolution of city sizes have not
been addressed at all in the presentation of Simon's model. Let us
point out three of them. In the first place, we have avoided a
detailed description of death and emigration events. We have in fact
assumed that the growth of the total population in the urban system
is monotonous, the only effect of mortality and emigration being a
lengthening of the duration of the evolution step (cf. Section
\ref{Simonmod}). This excludes the possibility that the population
might temporarily decrease --a necessary event if one aims at
describing, for instance, the eventual disappearance of cities. As
discussed elsewhere \cite{MZJTB}, a separate consideration of
mortality and/or emigration implies a change in the Zipf exponent
predicted by the model. Secondly, we have not taken into account the
possibility of migration flows inside the urban system, between its
cities. One can see that a purely multiplicative migration mechanism
would exchange population between cities without modifying the city
size distribution. On the other hand, additive and nonlinear
mechanisms would imply a change in the distribution. This belongs to
the class of open problems mentioned at the end of Section
\ref{destin}. Third, we have ignored possible events of coalescence
of cities which, as indicated in Section \ref{z}, shape many modern
urban systems. A particularly interesting open problem related to
such events regards the persistence of Zipf's law beyond the
formation of urban agglomerations. A model for this persistence may
shed light on the statistics of the coalescence process itself.

Finally, it is obvious that we have made no attempt to produce a
quantitative fitting of real data from city size distributions with
Simon's model or any of its extensions. On the other hand, very good
fittings have been reported for distributions of word frequencies in
language \cite{MM1,MM2}, musical notes in Western compositions
\cite{musZ}, and surname abundance\cite{MZJTB,Man:03}, all of which
share the dynamical basis of multiplicative processes. It would be
nice if this work elicits similar initiatives in the statistical
study of urban systems.

\section*{Acknowledgments}

Fruitful collaboration with Susanna Manrubia and  Marcelo Montemurro
is warmly acknowledged.


\begin{thebibliography}{0}

\bibitem{Gabaix2}Gabaix, X., Zipf's law for cities: An explanation,
{\it Quart. J. Econ.} {\bf  114}, 739--767 (1999).

\bibitem{Gabaix1}Gabaix, X. and  Ioannides, Y., The evolution of city size
distributions, in {\it Handbook of Urban and Regional Economics,
Vol. 4}, Henderson, V. and  Thisse, J. F., eds. (North-Holland,
Amsterdam, 2004).

\bibitem{Gibrat}Gibrat, R., {\it Les in\'egalit\'es \'economiques} (Librairie du
Recueil Sirey, Paris, 1931).

\bibitem{Man:03}Manrubia, S. C.,  Derrida, B. and  Zanette, D. H., Genealogy in the
era of genomics, {\it Am. Sci.} { \bf 91}, 158--165 (2003).

\bibitem{reset}Manrubia, S. C.,  and  Zanette, D. H., Stochastic multiplicative
processes with reset events, {\it Phys. Rev. } {\bf E59}, 4945--4948
(1999).

\bibitem{MZJTB} Manrubia, S. C.  and  Zanette, D. H., At the boundary between
biological and cultural evolution: The origin of surname
distributions, {\it J. Theor. Biol. } {\bf 216}, 461--477 (2002).

\bibitem{MM1}Montemurro, M. A. and  Zanette, D. H., New perspectives on
Zipf's law in linguistics: From single texts to large corpora, {\it
Glottometrics} {\bf 4}, 86--98 (2002).

\bibitem{Portu}Portugali, J., {\it Self-Organization and the City}
(Springer, Berlin, 2000).

\bibitem{Simon}Simon, H. A., On a class of skew distribution functions,
{\it Biometrika} {\bf 42}, 425--440 (1955).

\bibitem{Sornette1}Sornette, D.,  Multiplicative processes and power laws,
{\it Phys. Rev. } {\bf E57}, 4811--4813 (1998), and references
therein.

\bibitem{Sornette2}Sornette, D., {\it Critical Phenomena in Natural
Sciences. Chaos, Fractals, Selforganization and Disorder: Concepts
and Tools} (Springer, Berlin, 2000), and references therein.

\bibitem{lang}Stauffer, D. and Schulze, C., Microscopic and macroscopic simulation
of competition between languages, {\it Phys. Life Rev.} {\bf 2},
89--116 (2005).

\bibitem{Yule}Yule, G. U., A mathematical theory of evolution, based on the
conclusions of Dr. J. C. Willis, {\it Proc. R. Soc. London} {\bf
B213}, 21--87 (1924).

\bibitem{nos}Zanette, D. H.  and  Manrubia, S. C.,  Role of intermittency in
urban development: A model of large-scale city formation, {\it Phys.
Rev. Lett.} {\bf 79}, 523-526 (1997).

\bibitem{PhysA}Zanette, D. H.  and  Manrubia, S. C., Vertical transmission of
culture and the distribution of family names, {\it Physica} {\bf
A295}, 1--8 (2001).

\bibitem{MM2}Zanette,  D. H. and  Montemurro, M. A., Dynamics of
text generation with realistic Zipf's distribution, {\it J. Quant.
Linguistics} {\bf 12}, 29--40 (2005).

\bibitem{musZ}Zanette, D. H., Zipf's law and the creation of
musical context, {\it Musicae Scientiae} {\bf 10}, 3--18 (2006).

\bibitem{Zipf1}Zipf, G. K., {\it The Psycho-Biology of Language}
(Houghton-Mifflin, Boston, 1935).

\bibitem{Z0}Zipf, G. K., {\it Human Behaviour and the Principle of
Least-Effort} (Addison-Wesley, Cambridge, 1949).

\end{thebibliography}
\end{document}